\documentclass[a4paper]{jpconf}
\usepackage{graphicx}

\newcommand{\nggn}{(n,$\gamma$)-($\gamma$,n)}

\begin{document}
\title{Nucleosynthesis in neutrino-driven winds: influence of
  the nuclear physics input}

\author{Almudena Arcones \& Gabriel Mart\'inez-Pinedo}

\address{ Institut f\"ur Kernphysik,TU~Darmstadt, Schlossgartenstr.~9, D-64289 Darmstadt, Germany \\
  GSI Helmholtzzentrum f\"ur Schwerionenforschung GmbH, Planckstr.~1,D-64291 Darmstadt, Germany}

\ead{a.arcones@gsi.de}

\begin{abstract}
  We have performed hydrodynamical simulations of the long-time
  evolution of proto-neutron stars to study the nucleosynthesis using
  the resulting wind trajectories. Although the conditions found in
  the present wind models are not favourable for the production of
  heavy elements, a small enhancement of the entropy results in the
  production of r-process elements with A~$\approx$~195. This allows
  us to explore the sensitivity of their production to the
  hydrodynamical evolution (wind termination shock) and nuclear
  physics input used.
\end{abstract}

\section{Introduction}
Half of the heavy elements are produced by rapid neutron capture
(known as r-process). In the last years different scenarios have been
proposed, but many questions remain still open (see
\cite{arnould.goriely.takahashi:2007} for a review). There are two
conditions for a successful r-process: 1) the number of neutrons per
seed nuclei has to be high ($\approx 100$).  2) Once the astrophysical
site has been identified, nucleosynthesis calculations based on
simulations have to reproduce observations. In order to get the ratio
between two elements as observed in ultra metal poor stars, in
addition to high neutron-to-seed ratio, one needs to find the right
evolution of matter along the NZ plane, even after r-process freezes
out and matter moves back to stability. The question then emerges:
where in the Universe can such conditions be realized?

Our results presented here show that the final abundances depend
strongly on dynamical evolution and on the variation of the nuclear
physics input.  In particular, we have explored the influence of
various theoretical mass models with consistent sets of neutron
capture rates.

\section{Neutrino-driven wind}
There are several possible scenarios where r-process takes place
\cite{arnould.goriely.takahashi:2007}. Based on galactic chemical
evolution \cite{Ishimaru.Wanajo:1999} core-collapse supernovae and the
subsequent neutrino-driven winds remain the most promising
scenario. Therefore, we concentrate our study on neutrino-driven winds
\cite{arcones.janka.scheck:2007}, which are baryonic outflows from
proto-neutron stars formed in core-collapse supernova explosions. This
supersonic outflow expands through slow-moving matter, which was
ejected at the beginning of the explosion and leads to a wind
termination shock or reverse shock. This shock has a great influence
on the dynamical evolution of the outflow, and therefore on the
nucleosynthesis.

The conditions for r-process in neutrino-driven winds are well
understood \cite{Otsuki.Tagoshi.ea:2000} but requires extreme high
entropy, fast expansion or very low electron fraction. In recent
simulations \cite{arcones.janka.scheck:2007}, that we use for our
study, the entropies are too low or the electron fraction too high to
get a high enough neutron-to-seed ratio. Therefore it is still an open
question whether there is some physical aspects missing in the
long-time evolution of the ejecta or if the r-process takes place in
some other astrophysical site. More investigations of the
neutrino-driven wind and other astrophysical environments are
necessary to understand where heavy elements are created. However, the
available simulations are a useful basis to explore the influence of
the nuclear physics input on nucleosynthesis, when the entropy is
increased by a small factor. This simulates a dynamical evolution of
the matter that will be similar also to other astrophysical
environments where the r-process may occur.

\section{Nucleosynthesis studies}
We take the temperature and density evolution from trajectories of
hydrodynamical simulations \cite{arcones.janka.scheck:2007} as input
to the nucleosynthesis network combined with a given electron
fraction.  In addition, the density is reduced by a factor of two in
order to increase the entropy. As discussed, this allows us to study
the later r-process phase.

\begin{figure}[h]
\begin{minipage}[b]{37pc}
  \includegraphics[width=22pc]{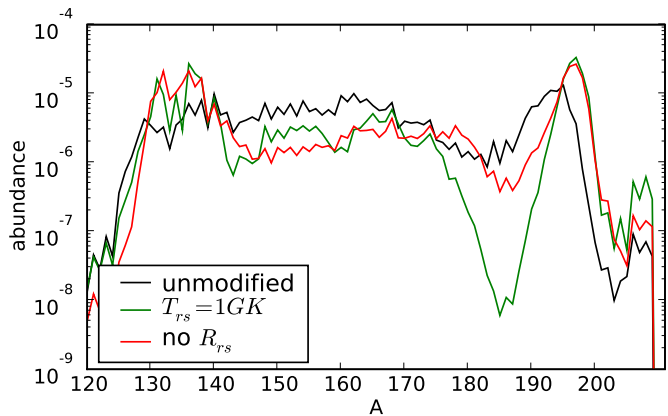}
  \caption{\label{fig:abund_rs} Final abundances for the same
    trajectory (i.e., same initial neutron-to-seed ratio) but
    different position of the reverse shock. For the black line, the
    position of the reverse shock is not changed from the simulation
    \cite{arcones.janka.scheck:2007}. For the green line, the reverse
    shock has been moved to a temperature of 1~GK. The red line
    represents a case without reverse shock, where we have taken the
    expansion to be adiabatic at the temperature of the former reverse
    shock. In all cases the entropy has been increased by a factor of
    two compared to the value from the simulation
    \cite{arcones.janka.scheck:2007}. The mass model used here is
    ETFSI-Q \cite{etfsiq}. Other mass models will be explored in
    Fig.~\ref{fig:abund_mm}.}
\end{minipage}
\end{figure}

\subsection{Dependence on late-time dynamical evolution and the
  reverse shock}
The r-process starts when the ejecta reach temperatures of around 1~GK
and is very sensitive to its subsequent evolution, that depends on the
interaction of the wind with the supernova ejecta. This interaction
produces a wind termination shock which is not a steady-state
phenomenon, therefore, it can only be studied within a full
hydrodynamical simulation.  However, there are some possibilities to
parametrize the behaviour of the reverse shock using an artificial
outer boundary with constant pressure
\cite{Sumiyoshi00,Terasawa.Sumiyoshi.ea:2002}, temperature
\cite{Wanajo.Itoh.ea:2002,Wanajo:2007}, or density
\cite{Kuroda.Wanajo.Nomoto:2008,Panov.Janka:2009}.  We have used data
directly from our simulations \cite{arcones.janka.scheck:2007} and
subsequently changed the position of the reverse shock in a consistent
way (for details see Ref.~\cite{arcones.prep}).

We find that, when the wind termination shock occurs at high
temperature ($\sim$~1~GK), the r-process takes place in \nggn
~equilibrium, as in the classical r-process. In this case the neutron
separation energies determine the abundances along an isotopic
chain. On the other hand, when the reverse shock is at low
temperatures ($<$~0.5~GK) photo-dissociation becomes negligible and
there is a competition between neutron capture and beta decay.
Consequently the relevant nuclear physics input will depend on the
dynamical evolution of the outflow. This translates into different
r-process paths and differences in the final abundances, as shown in
Fig.~\ref{fig:abund_rs}.

\subsection{Dependence on mass models}
\begin{figure}[h]
\includegraphics[width=22pc]{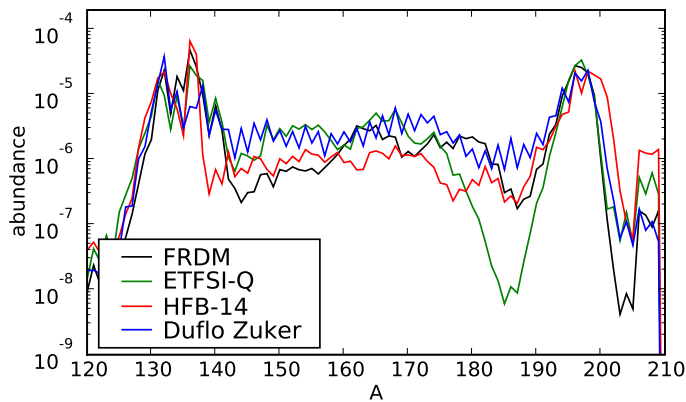}\hspace{0.5pc}%
\begin{minipage}[b]{12pc}\caption{\label{fig:abund_mm} Final
    abundances obtained from different mass models. All cases are
    calculated with the same trajectory which reaches \nggn
    ~equilibrium, because the reverse shock is at 1~GK.}\hspace{0.5pc}
\end{minipage}
\end{figure}

Here we describe only the case where the evolution takes place at high
temperatures and therefore in \nggn ~equilibrium. Other cases and more
details can be found in Ref.~\cite{arcones.prep}. In
Fig.~\ref{fig:abund_mm} we present the final abundances obtained for
the same trajectory but different mass models. We find that the region
before the third peak is remarkable different but also the abundances
between peaks vary considerably.

\begin{figure}[h]
  \begin{minipage}{18pc}
    \includegraphics[width=18pc]{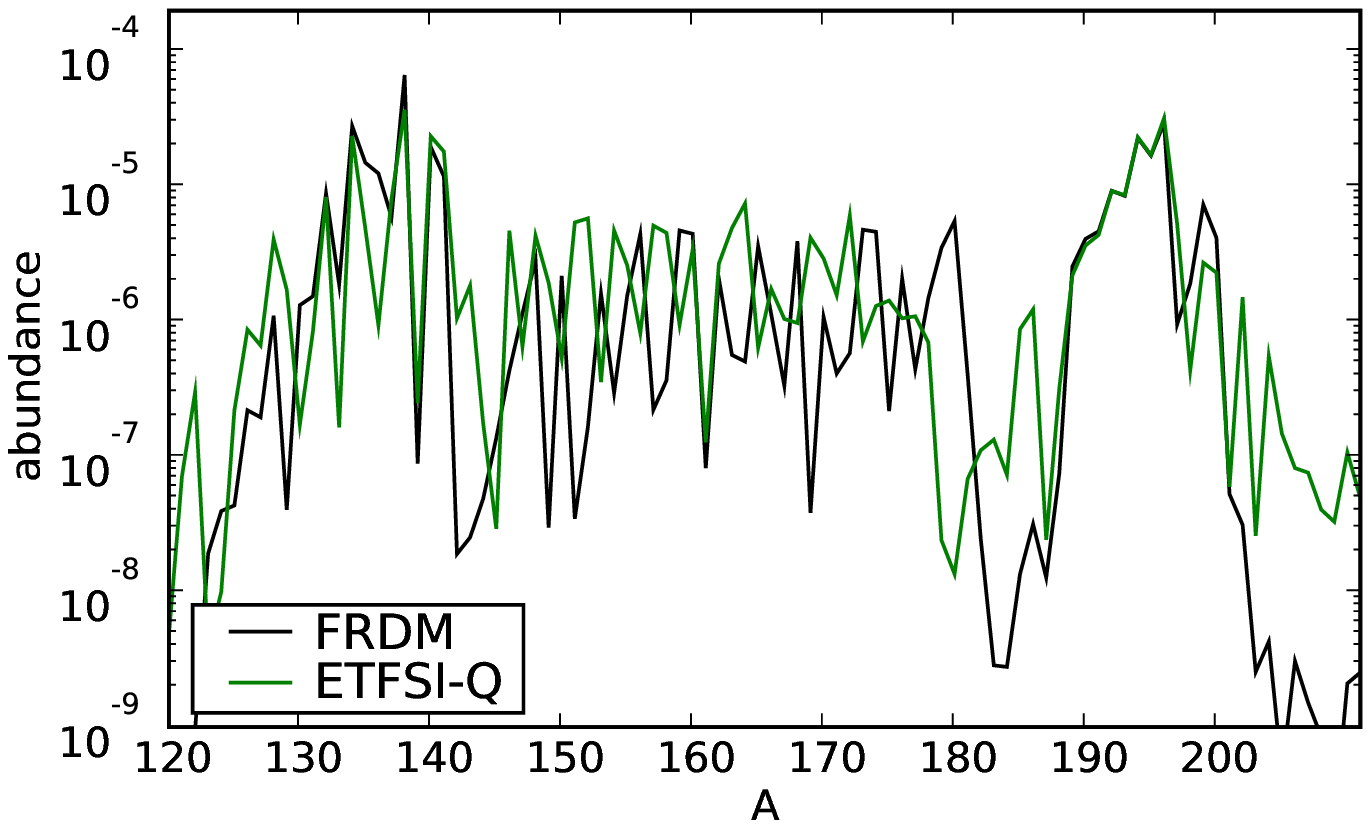}
    \caption{\label{fig:nseed1_abund} Abundances for
      $Y_n/Y_{seed}=1$.}
  \end{minipage}\hspace{2pc}%
  \begin{minipage}{18pc}
    \includegraphics[width=18pc]{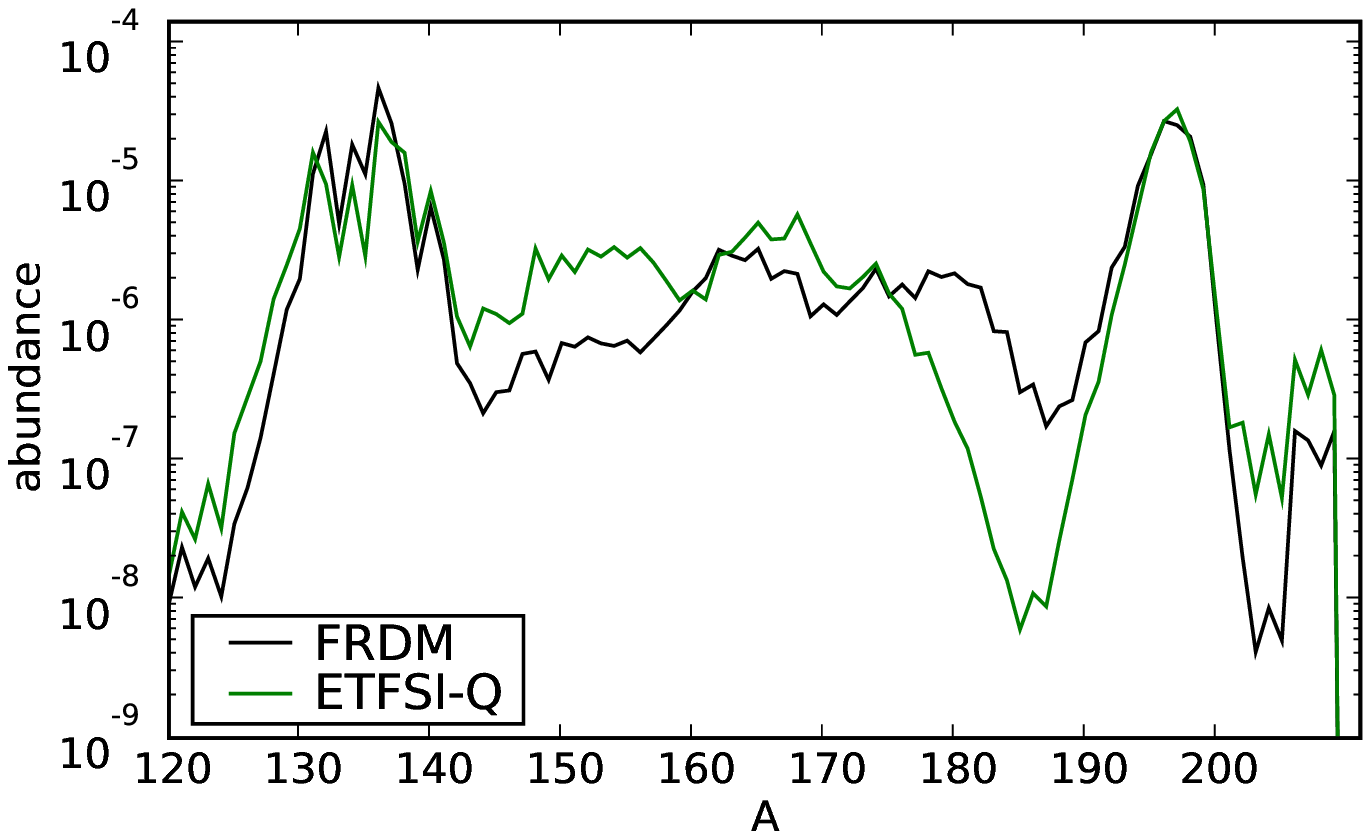}
    \caption{\label{fig:final_abund} Final abundances after all
      decays.}
  \end{minipage} 
\end{figure}

In order to assess the impact of nuclear masses, we compare the
abundances obtained from FRDM to ETFSI-Q at freeze-out
($Y_n/Y_{\mathrm{seed}}=1$) and when all decays have occurred.  In
Fig.~\ref{fig:nseed1_abund} we observe remarkable odd-even effects
following the behavior of the neutron separation energies. However,
the final abundances in Fig.~\ref{fig:final_abund} are smoother
similar to solar abundances.

In the long-time evolution there is a competition between beta decay
and neutron capture and we have found that neutron captures still play
an important role when matter moves back to stability, even when
neutron densities and neutron-to-seed ratios are low ($N_n\approx
10^{17} \mathrm{cm}^{-3}$ and $Y_n/Y_{\mathrm{seed}}\approx
10^{-5}$). Neutron captures can fill holes, move peaks to higher mass
number and reduce odd-even effects in the abundances.  Moreover, the
masses also enter in the neutron capture cross sections, and this can
explain the differences in Fig.~\ref{fig:final_abund} between the two
mass models.

\section{Conclusions}
The late-time evolution of the ejecta, also after freeze-out of the
r-process, is very important to determine details in the final
abundances. Therefore, we performed nucleosynthesis studies in
neutrino-driven winds by means of long-time hydrodynamical simulations
of core-collapse supernova explosions. The conditions found in the
simulations (low wind entropies and/or high electron fraction) do not
allow the formation of heavy elements. However, an artificial increase
of the entropy by a factor around two is enough to reach A=195 and
allow us to explore the sensitivity of the wind termination shock and
the nuclear physics input.

Depending on whether the evolution takes place at high or low
temperatures the relevant nuclear physics input will change. When
\nggn equilibrium is valid the masses and beta decays determine the
abundances. In the low temperature case, beta decays and neutron
capture are crucial. In both cases, the final abundance distribution
is determined by a competition between neutron capture and beta
decays, when matter moves back to stability.

\subsection{Acknowledgments}
A.~Arcones acknowledges support by the Deutsche
Forschungsgemeinschaft through contract SFB 634.

\section*{References}
\providecommand{\newblock}{}


\end{document}